\documentclass[12pt]{article}
\usepackage{xspace}
\usepackage{rotating,amsmath,amssymb}
\usepackage{epsfig}\parskip 5pt plus 1pt
\newcommand{\be}{\begin{equation}}
\newcommand{\ee}{\end{equation}}
\newcommand{\bea}{\begin{eqnarray}}
\newcommand{\eea}{\end{eqnarray}}

\def\simge{\mathrel{%
   \rlap{\raise 0.511ex \hbox{$>$}}{\lower 0.511ex \hbox{$\sim$}}}}
\def\simle{\mathrel{
   \rlap{\raise 0.511ex \hbox{$<$}}{\lower 0.511ex \hbox{$\sim$}}}}

\def\nue{\ensuremath{\nu_{e}}\xspace}

\def\numu{\ensuremath{\nu_{\mu}}\xspace}

\def\Li{^6{\mathrm{Li}}}
\def\Li8{\ensuremath{^8{\mathrm{Li}}}}
\def\B8{\ensuremath{^8{\mathrm{B}}}}

\def\anumu{\overline{{\mathrm\nu}}_{\mathrm \mu}}

\newcommand{\numunue}{\ensuremath{\nu_\mu \rightarrow \nu_e}\xspace}

\newcommand{\kthree}{\ensuremath{ K_{e3}}\xspace}

\begin{document}
\thispagestyle{empty}
\vspace*{1cm}
\begin{center}
{\Large{\bf Electron neutrino tagging through tertiary lepton detection} }\\
\vspace{.5cm} 

L.~Ludovici$^{\rm a}$ and
F.~Terranova$^{\rm b}$ \\
\vspace*{1cm}
$^{\rm a}$ 
I.N.F.N., Sezione di Roma Sapienza, Rome, Italy \\
$^{\rm b}$ I.N.F.N., Laboratori Nazionali di Frascati,
Frascati (Rome), Italy \\

\end{center}

\vspace{.3cm}
\begin{abstract}
\noindent
We discuss an experimental technique aimed at tagging
electron neutrinos in multi-GeV artificial sources on an
event-by-event basis. It exploits in a novel manner calorimetric and
tracking technologies developed in the framework of the LHC
experiments and of rare kaon decay searches. The setup is suited for
slow-extraction, moderate power beams and it is based on an
instrumented decay tunnel equipped with tagging units that intercept
secondary and tertiary leptons from the bulk of undecayed $\pi^+$ and
protons. We show that the taggers are able to reduce the \nue
contamination originating from \kthree decays by about one order of
magnitude. Only a limited suppression ($\sim$60\%) is achieved for
\nue produced by the decay-in-flight of muons; for low beam
powers, similar performance as for \kthree can be reached
supplementing the tagging system with an instrumented beam dump.
\end{abstract}

\vspace*{\stretch{2}}
\begin{flushleft}
  \vskip 2cm
{ PACS: 14.60.Pq, 29.40.Vj, 95.55.Vj} 
\end{flushleft}

\newpage

\section{Introduction}
\label{introduction}

The identification of the initial flavour of neutrinos produced by
artificial sources through the detection of the associated lepton
(``neutrino tagging'') is a possibility that has been envisaged many
decades ago~\cite{hand,pontecorvo}. Its realization, however, must
overcome major experimental challenges and, in spite of numerous
proposals~\cite{Nedyalkov:1981as,Nodulman:1982ua,ammonosov,bernstein,ludovici},
a facility operating with \nue or \numu neutrinos tagged on an
event-by-event basis is still to come. In a tagged neutrino facility,
a precise knowledge of the neutrino flavour at the source can be
achieved identifying the associated lepton in coincidence with the
occurrence of a neutrino interaction at the far detector. As it will
be shown in the following, the exploitation of this correlation
requires time resolutions below 1~ns both for the taggers and for the
neutrino detectors. In the past, two approaches have been pursued. The
former - dating back to 1969~\cite{hand} - is targeted to the
identification of $\nu_\mu$ from kaon decay: it takes advantage of the
large difference in Q-value between $\pi$ and $K$ decays to isolate
leptons from kaons without intercepting muons from $\pi$ or the
undecayed parent mesons. The second approach focuses on the
identification of positrons in order to either select a pure \nue beam
for physics measurements~\cite{bernstein} or to veto the \nue
contamination in \numu beams from $\pi$ decay-in-flight
(``anti-tagging''~\cite{ludovici}). The technique that we discuss in
this paper combines both approaches in a novel manner; moreover, it
takes advantage of the outstanding progresses in high-rate
radiation-hard detectors that have been achieved for the calorimetry
in the forward region of LHC experiments and for the study of very
rare kaon decays. As in~\cite{ludovici}, the tagging setup discussed
hereafter is especially suited to suppress the \nue contamination in
\numu beams from the decay in flight of multi-GeV pions, i.e. to
reduce the intrinsic contamination of beam-related \nue events at
experiments seeking for \numunue transitions. It can also be employed
to reduce the systematic error in the knowledge of the flavor
contamination at source or to select a pure \nue subsample for physics
studies. The physics reach of such experiments ranges from the study
of anomalous short baseline 
oscillations~\cite{miniboone,microboone,stancu,rubbia} 
and low energy cross sections~\cite{Nakaya:2008zz},
which employs low power proton beams on solid targets, up to ambitious
``Superbeam'' facilities~\cite{T2K,NOVA,frejus,WBB} utilizing MW-class
proton drivers to address the study of subdominant \numunue
transitions at the atmospheric scale. All these experiments are
limited by systematic errors~\cite{miniboone_lowe,mezzetto_sys} mainly
arising from the finite purity of the neutrino source: as a
consequence, in the last decade the development of novel facilities
designated to overcome the purity constraints of $\pi$-based beams has
been at focus of intense R\&D efforts~\cite{ISS_reports}. In this
paper, the tagging principle and the conceptual design of a facility
aimed at a substantial reduction of the \nue contamination is
discussed in Sec.~\ref{sec:setup}. Simulation and performance for
background rejection in a setup ``scraping'' the secondary beam until
the hadron dump is studied in Sec.~\ref{sec:performance}.  Finally,
the special case of an additional instrumented beam dump at the end of
the decay tunnel is discussed in Sec.~\ref{sec:beam_dump}.

\section{Tagging of electron neutrinos}
\label{sec:setup}

\begin{figure}
\centering
\epsfig{file=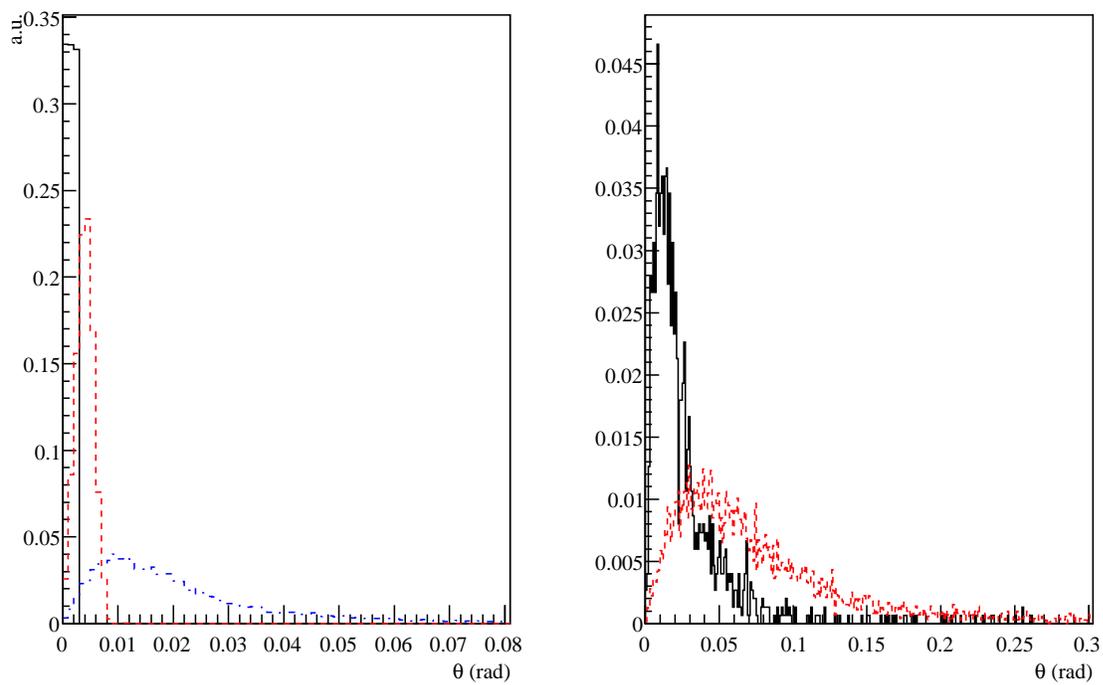,width=\textwidth}
\caption{Left: $\theta$ angle distribution between the propagation
direction of pions (black solid line), muons (red dashed) and
$\nu_\mu$ (blue dot-dashed) and the axis of the decay tunnel.  Right:
$\theta$ distribution for positrons resulting from DIF (black
solid line) and \kthree decays (red dashed line).  }
\label{fig:angles}
\end{figure}

Artificial sources of \numu at energies larger than $\sim$100~MeV can
be produced by the two-body decay in flight of pions $\pi^+
\rightarrow \mu^+ \nu_\mu $, which in turn are created from protons
impinging on thick targets~\cite{kopp}. The source also contains \numu
originating from two-body and three-body decays of charged and neutral
kaons. However, it is intrinsically polluted by \nue originating from
three-body decays of $K^+$ ($K_{e3}$: $K^+ \rightarrow \pi^0 \nue
e^+$), from the decay-in-flight of secondary muons along decay
tunnel (DIF: $\pi^+ \rightarrow \mu^+ \nu_\mu \rightarrow e^+ \nue \numu
\anumu$) and from the decays of neutral kaons. Tuning of the pion
momentum selected by magnetic lenses after the primary target, of the
transfer line up to the decay tunnel and, finally, of the length and
radius of the tunnel itself help in reducing the ratio \nue/\numu
below 1\%, although a \nue contaminations in the 1-0.1\% range is
unavoidable in any realistic configuration. In a very broad neutrino
energy range, i.e. from sub-GeV~\cite{miniboone_flux} up to tens of
GeV~\cite{flux_CNGS}, the \nue contamination is dominated by $\pi^+$
DIF and by $K_{e3}$, with the addition of a minor contribution from
semileptonic decay of $K^0_L$ ($K^0_L \rightarrow \pi^- e^+ \nue$)
and charged kaons decaying before the bending magnets.
All these decays produce positrons in the final state, whose spectral
distribution follows from three-body kinematics and from the specific
Q-value of the reaction. In the forward region along the decay tunnel,
positrons are swamped by the bulk of undecayed hadrons, by muons
resulting from the two-body decay of $\pi^+$ and by secondary protons
within the acceptance of the focusing system. Fig.~\ref{fig:angles}
shows the polar angle distribution $\theta$ of positrons, muons and
electrons with respect to the axis of the decay tunnel for a specific
beam configuration (``benchmark beamline'' - see
Sec.~\ref{sec:performance}) resulting from the decay in flight of
$\pi^+$ and $K^+$ with 8.5~GeV mean energy. At the entrance of the
decay tunnel, mesons and protons are assumed to be uniformly distributed in a $10
\times 10$~cm$^2$ area with a polar angle smaller than 3~mrad (black solid
curve in Fig.~\ref{fig:angles}-left). On top of the beam divergence,
positrons from $K_{e3}$ (Fig.~\ref{fig:angles}-right) show a large
intrinsic divergence due to three-body kinematics, with a mean
$\theta$ value of 88~mrad. The mean polar angle of muon-neutrinos
produced by the source is of the order of 27 mrad (blue dot-dashed line in
Fig.~\ref{fig:angles}-left), i.e. it has a value intermediate between
the $\pi^+$ and positron beam divergence. Similarly, the $\theta$
distribution of positrons from DIF has a mean value of 28~mrad (black
solid line of Fig.~\ref{fig:angles}-right).

\begin{figure}
\centering
\epsfig{file=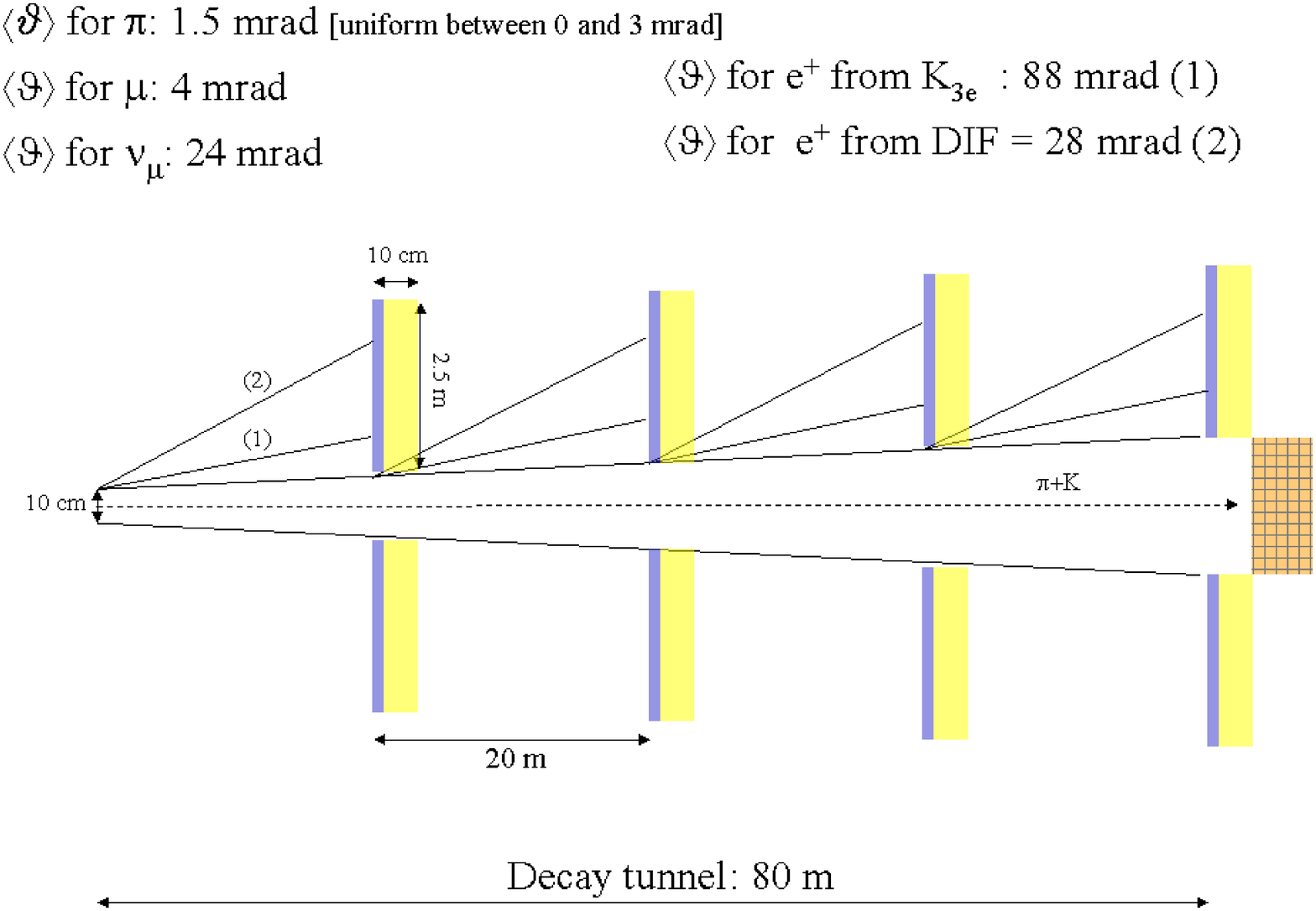,width=0.8\textwidth}
\caption{Schematics of the tagging setup (lateral view, not in
scale). The blue\&yellow (solid) boxes indicate the position and size of the tagging
modules (tracker and calorimeter) along the 80~m decay tunnel. The lines labeled (1) and
(2) show the average $\theta$ angles for positrons from \kthree and
DIF, respectively. The orange (gridded) box represents the instrumented dump
(Sec.~\ref{sec:beam_dump}).}
\label{fig:setup}
\end{figure}

On the other hand, the polar angle of the accompany muons in $\pi^+
\rightarrow \mu^+ \nu_\mu$ along the decay tunnel is extremely small
(red dashed in Fig.~\ref{fig:angles}-left) and, in fact, comparable
with the beam divergence of the parent $\pi^+$. This muon focusing
effect is due to 2-body kinematics and to the fact that the rest mass
of the muon is comparable with the pion rest mass. The emission angle
$\theta$ of the muon is, therefore,
\be
\tan \theta = \frac{\sin \theta^*}{\gamma (1/\beta^* + \cos \theta^*) }
\ee
$\gamma$ being the Lorentz boost of the parent pion ($\beta \simeq 1$)
in the laboratory frame, while $\theta^*$ and $\beta^* =
(m_\pi^2-m_\mu^2)/(m_\pi^2+m_\mu^2) \simeq 0.26$ are the emission
angle and the muon velocity in the pion rest frame, respectively. On
the contrary, $\beta^* = 1$ for neutrinos; hence, the emission angle
in the laboratory frame at large $\theta^*$ ($\cos \theta^* \simeq 0$)
is much wider than for $\mu^+$.  The above considerations point toward
a tagging setup designed to perform a destructive (calorimetric)
measurement of the positrons and intercepting secondary particles
emerging from the primary pion beam up to the end of the decay
tunnel. The setup is shown schematically in Fig.~\ref{fig:setup}. It
consists of a set of cylindrical e.m. calorimeters with a geometry and
readout similar to the ATLAS forward calorimeter (FCAL~\cite{atlas})
but with much shorter length (10~$X_0$). The FCAL is a liquid Argon
calorimeter designed to operate in a high radiation density environment. It
ensures radiation hardness up to 0.5 GRad/y, fast response to cope
with the 25~ns beam crossing of the LHC and reduced sensitivity to
event pile-up thanks to a very small drift length (0.27~mm). Such
small gap allows for full particle drift in 61~ns and, therefore,
relieves the detector of the problem of ion
build-up~\cite{rutherfoord}. Unlike ATLAS, the modules considered here
are built with inner radii of variable size, so that all primary
mesons and most secondary muons reach the beam dump without intercepting
the calorimeters. Differently from early
proposals~\cite{Nedyalkov:1981as,ludovici} there is no material
installed along the trajectory of the undecayed pions up to the end of
the decay tunnel; therefore, irrespective to the tagging performance,
the flux of \numu at the far detector remains unchanged after the
installation of the modules.  In front of each module, a high speed
tracker of granularity comparable to the FCAL ($\sim 1$~cm) is
envisaged. The tracker provides an absolute time-stamp of the incoming
candidate positron with a required precision of $\sim$1~ns (see
Sec.~\ref{sec:performance}) and vetoes neutral energy deposits due to
energetic photons from $\pi^0$ or hard bremsstrahlung. In the
occurrence of a neutrino interaction at the far location, a time
coincidence with a charged particle in the tracker is sought for, once
accounting for the propagation delay due to the source to the detector
distance. In addition, an electromagnetic deposit beyond a given
threshold is required in the FCAL area adjacent to the tracker hits
within a time window comparable with the drift time of the
calorimeter. The presence of such energy deposit indicates the
simultaneous production of a neutrino with a positron, tagging the
event at the far location as a ``\nue at source''. The finite
efficiency and geometrical acceptance of the modules (e.g. due to
positron produced along the pion flight direction) limits the
capability to veto \nue (Sec.~\ref{sec:performance_eff}). Fake vetoes
due to accidental overlaps of muons with photons, e.m. shower leakage
along the modules and $\mu \rightarrow e$ misidentification induces
additional dead-time, causing a drop in statistics
(Sec.~\ref{sec:accidentals}).  As discussed in
Secs.~\ref{sec:performance} and \ref{sec:beam_dump}, in order to
achieve a \nue suppression rate of about one order of magnitude and a
sizable detector livetime, the tracker must be able to operate with
rates up to $\sim$200~kHz/cm$^2$ ($\sim$20~MHz/cm$^2$ in the proximity of the beam
dump) and a time resolution better than 1~ns. Such requirements point
toward the use of fast semiconductor detectors\footnote{Other options
based on scintillators or gaseous detectors can be envisaged,
especially in the areas far from the secondary meson beam, where the
requirements on the particle rate can be relaxed below 10~kHz/cm$^2$
(see Sec.~\ref{sec:performance}).}: they share with collider physics
applications the constraints coming from radiation
hardness~\cite{radhard} and with rare kaon decay physics applications
the need for small material budget and sub-ns time
resolution~\cite{na62}. However, the granularity requirements are
highly reduced (1~cm versus a few hundreds $\mu m$) and the
constraints on the time resolution can be relaxed by about one order
of magnitude (1~ns versus 100~ps); still, fast trackers for neutrino
tagging applications represent a technological challenge due to the
large area needed for full coverage of the calorimeters in the region
where particle rates are high ($\sim$10~m$^2$ for rates higher than
than $10$~kHz/cm$^2$). The requirement of a time resolution better
than 1~ns poses strict constraints to the neutrino detector, as well,
thus narrowing the choice of technologies that can be
employed. Cherenkov detectors, in particular, offer the advantage of
extremely fast response at the expense of reduced light yield with
respect to scintillators or gaseous detector. In recent years, water
Cherenkov neutrino detectors have achieved resolutions below 1~ns for
masses up to 50~kton~\cite{SK_timing}, although fast triggering from
liquid and solid scintillators, from UV light in liquefied noble gases
and from gaseous detectors still represent viable alternatives.
Finally, it is worth noticing that an absolute time calibration
between the instrumented decay tunnel and the neutrino detector must
match the above-mentioned resolutions. For short baseline experiments,
it can easily be achieved with atomic clocks, while the technique
currently exploited by long baseline experiments and based on the
Global Positioning System~\cite{gps} does not fulfill this constraint
($\Delta t \simeq 10$~ns), so that a direct resynchronization of the
clocks would be periodically needed.


\section{Performance of the tagging system}
\label{sec:performance}

\subsection{Event selection and tagging efficiency}
\label{sec:performance_eff}

\begin{figure}
\centering
\epsfig{file=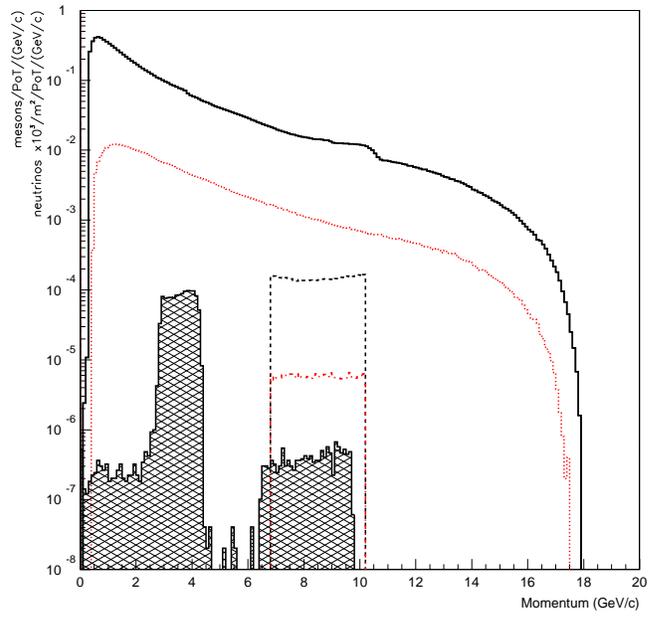,width=0.6\textwidth}
\caption{Muon neutrino flux (hatched histogram) in a detector located 800m from
the target. Spectra of $\pi^+$ (black solid line) and $K^+$ (red dotted) 2~m downstream
the target. Spectra of $\pi^+$ (black dashed) and $K^+$ (red dot-dashed) selected
by the focusing and bending system.}
\label{fig:fluxes}
\end{figure}

The tagging concept outlined in Sec.~\ref{sec:setup} has been tested
against a specific beam configuration in order to quantify the
performance of the \nue suppression. As already mentioned, the two
most relevant contributions to the \nue beam contamination are due to
\kthree and DIF. The beam setup considered hereafter (``benchmark
beamline'') is similar to the one studied in \cite{ludovici} and it is
suited for short baseline \nue appearance searches, along the line of
\cite{miniboone,microboone,stancu,rubbia}. In fact, the beamline of
\cite{ludovici} has been considered as an upgrade of the I216
Proposal~\cite{I216} at the CERN PS; here, it is used as a benchmark
and the number of pions produced per extraction, together with the
proton mean power are considered as free parameters.  The neutrino
beam in this benchmark configuration is produced by a slow extraction
of protons (1~s) from a 19.2~GeV booster. In our
simulation~\cite{geant3} protons are dumped in a cylindrical beryllium
target (110~cm length and 3~mm diameter) producing secondary
particles, which are momentum-selected and transported up to the decay
tunnel by a magnetic focusing system.  The focusing system necessarily
relies on quadrupole-dipole magnets~\cite{kopp} due to the long
extraction spill and has not been simulated in details (halo muons and
off-momentum particles are, therefore, neglected). Similarly to
\cite{ludovici}, we assume the focusing system to have an angular
acceptance of 80~$\mu Sr$ and a momentum bite of $\pm20$\% around a
nominal momentum of 8.5~GeV.  All selected secondary particles are
focused at the entrance of the decay tunnel, uniformly distributed in
a 10$\times$10~cm$^2$ window. Here, in order to simulate the beam
divergence, the secondary particles are randomly distributed up to an
angle of 3~mrad with respect to the axis of the decay tunnel.  We
cross-checked our simulation with a Sanford-Wang \cite{sanfw}
parametrization of pion production data \cite{harpfw} and with $\pi^+$
and $K^+$ data taken with 19.2~GeV proton on beryllium \cite{allaby}.
The decay tunnel is 80~m long with a 2.5~m radius
(Fig.~\ref{fig:setup}).  We neglect here the production of $K_L^0$ and
$K^-$ at the target, whose \nue contribution at the far detector
depends on the details of the focusing system. With respect to
traditional neutrino beams, this contribution is further suppressed
due to the bending between the primary target and the decay tunnel; it
is, thus, negligible with respect to the \kthree and DIF contributions
even after tagging suppression. Assuming a primary beam intensity of
$2\times 10^{13}$ protons-on-target (pot) per
extraction~\cite{ludovici}, we evaluate from simulation a particle
density at the entrance of the decay tunnel of about $R_0\equiv
100$~MHz/cm$^2$ during the 1~s-long spill. It corresponds to
$N_\pi=10^{10}$ $\pi^+$ per extraction over the whole surface, with a
mean $K^+/\pi^+$ of 4.1\%.  Any increase of the neutrino flux achieved
by an increase of the time the booster is dedicated to the neutrino
beamline (``duty cycle'') does not affect the tagging
performance\footnote{For the benchmark beamline ($E_p=19.2$~GeV), a
1~s long extraction delivering $2 \times 10^{13}$ pot every 7~s
corresponds to a mean beam power of 9~kW and a duty cycle of
1/7=14\%.}.  However, a power increase obtained raising the number of
mesons $R_0$ per extraction, e.g. increasing the proton beam intensity
and energy or the acceptance of the focusing system, increases the
particle rate at the taggers and challenge the tagging performance.
For the benchmark beamline in our simulation, Fig.~\ref{fig:fluxes}
shows the spectra of all $\pi^+$ and $K^+$ produced and crossing a
circle of 1.4~m radius, located 2~m downstream the target and the
spectra of $\pi^+$ and $K^+$ accepted by the focusing and bending
system.  The neutrino flux in a far detector located 800~m from the
decay tunnel is also shown in Fig.~\ref{fig:fluxes}. The neutrino beam
is a narrow-band beam of $1.25 \times 10^{-7}$ $\nu / \mathrm{pot}
/\mathrm{m}^2$ at a far detector.  It gives $\sim 1.9 \times 10^{4}$
events every $10^{20}$ pot in a 1~kton detector.  The mean \numu
energy is 3.5~GeV and the \nue contamination is 0.1\%.

\begin{figure}
\centering
\epsfig{file=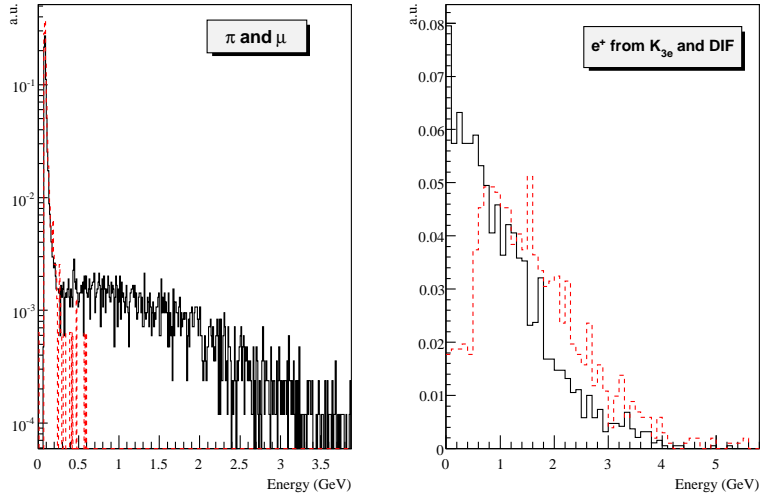,width=0.7\textwidth}
\caption{Left: energy at the tagger for undecayed $\pi^+$ (black solid
line) and $\mu^+$ (red dashed).  Right: energy at the tagger from
$e^+$ originating by \kthree (red dashed) and DIF (black solid) giving
a \nue at the far detector (``\nue at source'').  }
\label{fig:edep}
\end{figure}

Along the decay tunnel, four tagging stations have been
simulated. They are located $z=$20, 40, 60 and 80~m far from the
entrance of the tunnel ($z=0$). The stations have a cylindrical
geometry with 2.5~m outer radius and variable inner radii of 12, 17,
21 and 26~cm, corresponding to an angular opening of 6, 4.25, 3.5 and
3.25~mrad. Variable angular openings are employed since the source at $z=0$ is
non-pointlike: in this case most of undecayed primary $\pi^+$ and
$K^+$ reach the beam dump at $z=80$~m through the central holes of the
modules and only secondaries are scraped by the tagging units. Each
module has a material budget of 10~$X_0$ of lead\footnote{In fact, in
the special case of the Atlas FCAL, copper has been employed as
passive material.}, which has been simulated in GEANT4~\cite{geant4}. The
calorimeter response to energy deposit has not been simulated in
details: for e.m. deposits (positrons, electrons, photons) the
reconstructed energy is drawn from the deposited energy and smeared
according to the FCAL measured resolution~\cite{atlas_testbeam}. In
the present case, the energy resolution is dominated by the sampling
term, which is assumed to be 30\%/$\sqrt{E}$ with $E$ expressed in
GeV. Since the tagger thickness is just 33\% of the Pb interaction
length, pions deposit only a small fraction of their energy in the
tagger. Such deposit is mip-like in about 74\% of the case and it
exhibits strong energy fluctuations in the occurrence of hard hadronic
interactions. It is shown in Fig.~\ref{fig:edep}-left (black line) and
superimposed with the energy deposit of muons from $\pi^+$ decay (red
dashed line).

Fig.~\ref{fig:edep}-right shows the reconstructed energy of positrons
from \kthree and DIF, when a \nue reaches the far detector. The energy
has been smeared according to the FCAL resolution and accounting for
lateral leakage. For each event we require a \nue with energy larger
than 0.5~GeV hitting the far detector within its geometrical
acceptance. We assumed a source-to-detector distance of
800~m~\cite{ludovici}.  The detector surface in the plane
perpendicular to the neutrino beam is $10\times10$~m$^2$.  Lateral
leakage is marginal ($\sim$1\%) for positrons if all energy deposited
within a 2~cm radius ($R_c$) around the impact point is
collected. $R_c$ has been optimized empirically: larger collecting
radii are detrimental for $\pi$/e separation and for the effect of
event pile-up while radii much smaller than 2~cm reduce the visible
energy of the positrons, which gets closer to a mip-like deposit.

Finally, a ``\nue at source'' is defined as a charged particle
triggered by the tracker with an energy deposit in the calorimeter
greater than 300~MeV, in coincidence with a \nue interactions at the
far detector. Again, only neutrinos with energy larger than 0.5~GeV
are considered. The inefficiency of the tracker has been neglected
together with subdominant effects as the albedo resulting from
backscattered electrons or the different energy response for the
hadronic and e.m. component in the core ($R_c<2$~cm) of the $\pi^+$
shower.  Results in terms of \kthree and DIF suppression are listed in
Tab.~\ref{tab:result1}. The numbers in bold indicate the veto
efficiency for events giving a \nue at the far detector coming from
\kthree and DIF, respectively.  The tagging system, therefore, is
powerful in vetoing the \kthree contamination, while the performance
are poorer for DIF \nue. This is due both to the smaller angular
spread of DIF $e^+$ and to the different lifetime of kaons and
muons. Since the $\gamma c \tau$ of the kaon (64~m) is comparable with
the length of the decay tunnel, \nue from \kthree decay are produced
earlier than \nue from DIF, generating positrons that are intercepted
by the tagging modules. On the other hand, due to the longer pion and
muon lifetime ($\gamma_\pi c \tau_\pi \simeq 478$~m) most DIF occur in
the proximity of the beam dump and positrons impinge upon the dump at
a radius smaller than the inner radius of the last tagging unit. In
Tab.~\ref{tab:result1}, the improvement in vetoing the DIF is shown in
the third column, assuming that the last tagging unit has no inner
hole, i.e. instrumenting the beam dump (see Sec.~\ref{sec:beam_dump}),
and applying a tighter cut of 1~GeV in this area. As anticipated, the
improvement in the \kthree rejection is marginal (+3\%) while the
rejection of DIF increases substantially (+16\%).  The second and
fourth columns are computed without energy smearing (``NS'') and show
the impact of the sampling term of the e.m. calorimeter on the tagging
performance. Finally, it is worth noticing that, beyond tagging, the
instrumented decay tunnel allows for a precise measurement of the
secondary beam and, therefore, it contributes to reduce 
systematic errors on the $\nu$ flux and composition.

\begin{table*}
\centering
\begin{tabular} { c c c c c }
\hline
 & Scraping & Scraping (NS) & Dump & Dump (NS) \\
\hline
\kthree &  \bf{ 86.2 $\pm$ 0.3\%} & $88.5 \pm 0.3$\% & $89.1 \pm 0.3$\% & $91.4 \pm 0.3$\% \\   
DIF & {\bf 60.7 $\pm$ 0.7\%} & $64.4 \pm 0.7$\%  & $76.1 \pm 0.6$\% & $80.0 \pm 0.6$\% \\
\hline
\end{tabular}
\label{tab:result1}
\caption{Percentage of \nue from \kthree (first line) and DIF (second
line) at source vetoed by the tagging system. The first two columns
indicate the tagging efficiency assuming only the scraping of the
secondary beam as described in the text.  NS (no-smearing) shows
the efficiencies obtained considering an ideal e.m. calorimeter
(i.e. a negligible sampling term).  The last two columns show the
improvement gained introducing an instrumented beam-dump and applying
a tighter cut of 1~GeV on the reconstructed energy inside the dump
region (R$<$26~cm at the last tagging module). Errors are due to finite
MC statistics.}
\end{table*}

\subsection{Particle rate and accidental coincidences}
\label{sec:accidentals}

\begin{figure}
\centering
\epsfig{file=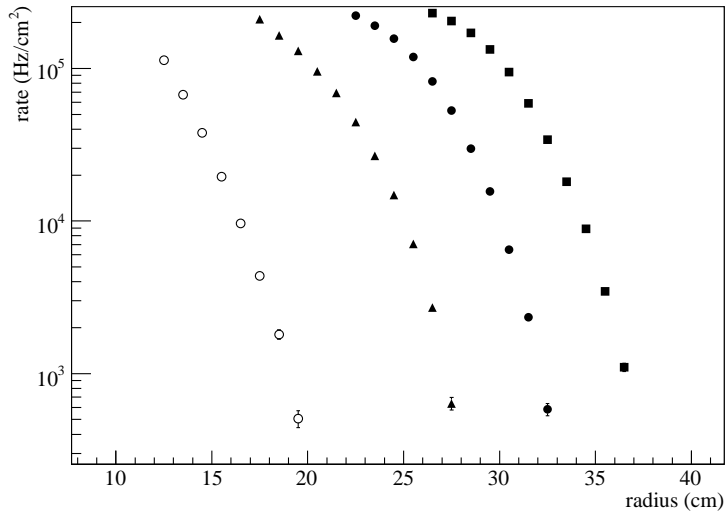,width=0.7\textwidth}
\caption{Rate at the first (20~m distance - empty dots), second (40~m,
triangles), third (60~m, full dots) and fourth (80~m, squares) tagging
module. The error bars (visible only at large radii) show the
errors due to finite MC statistics.}
\label{fig:rate}
\end{figure}

The particle rate at the tagging units is clearly dominated by the
2-body decay of the $\pi^+$ with $\mu^+$ impinging on the calorimeters
along the decay tunnel. In the beam dump (see Sec.~\ref{sec:beam_dump}), it
mainly depends on the rate of undecayed $\pi^+$, plus a small
correction due to $K^+$ and secondary protons that reaches the end of
the beamline.  Fig.~\ref{fig:rate} shows the average particle rate
during the 1~s extraction as a function of the tagger radius. As
anticipated, the rate at the taggers is dominated by muons, with peak
values of about 200~kHz/cm$^2$. As a consequence, the peak rate of
pile-up is
\be
PR = 2\times 10^5 \mathrm{cm}^{-2}\mathrm{s}^{-1}  S  \Delta T_{cal}
\ee
$S$ being the collection surface $\pi R_c^2 \simeq 12$~cm$^2$ and
$\Delta T_{cal}$ is the integration time of the detector. For the
above-mentioned FCAL, it corresponds to 61~ns if operated in
full-drift mode, but it drops below 25~ns in the standard LHC readout
configuration, which exploits the fast rise of the signal. Since
piling-up particles are mainly constituted by muons, PR up to $\sim 3$
are sustainable before the energy deposit reaches the threshold for
positron identification (300~MeV). For the benchmark beamline
$PR\simeq 0.06$, so that beam powers up to a few hundreds of kW would
be sustainable.  In particular, a pile-up of 3 is reached by a beam
power of 330~kW, assuming a duty cycle similar to the one of the
benchmark beamline (14\%).

Accidental coincidences are due to events classified as ``\nue at
source'' by the tagging system in the same time window ($\sim$1~ns) of
the neutrino interaction at the far location. Two classes of events
give contributions. 

Firstly, accidental tags can arise from
genuine $\nue$ at source in coincidence with \numu or \numunue
interactions in the detector. 
The probability of having a fake coincidence between an event at the
far detector and a ``\nue at source'' from the tagging system is given by
the rate of positrons impinging on the trackers times the squared sum
of the time resolution of the tracker and the neutrino detector
($\Delta t$).  The rate of accidentals is, therefore,
\be 
\left[ N_\pi f^\pi_{e^+} \epsilon_{DIF} + N_K f^K_{e^+} \epsilon_{\kthree} \right] \cdot \Delta t
\simeq 1.3 \times 10^{7} \cdot  \Delta t \ \ ;
\label{eq:acc}
\ee
$ \epsilon_{DIF} $ and $\epsilon_{\kthree}$ are the tagging efficiency
for DIF and \kthree positrons\footnote{Unlike Tab.\ref{tab:result1},
these efficiencies (63\% and 83\% respectively) are computed for all
positrons, irrespective of the energy and direction of the outcoming
\nue.}, $N_\pi \simeq 10^{10} s^{-1}$ the rate of pions at $z=0$ and $N_K \simeq 4.1 \times 10^{-2}
\cdot N_\pi$ the kaons at $z=0$. The fraction of $\pi^+$ ($K^+$)
giving a positron in the decay tunnel from DIF (\kthree) is labeled $
f^\pi_{e^+}$ ($f^K_{e^+}$) and, in the benchmark beamline is $\simeq
0.017$\% (3.4\%). In particular, $f^K_{e^+} = BR( \kthree ) (1-
e^{-\gamma_K c \tau_K / L})$, $L$ being the tunnel
length. Eq.~\ref{eq:acc} sets the scale of the time resolution needed
at the tracker and neutrino detector.  For $ \Delta t=1$~ns, the
fraction of \numunue interactions at the far detector wrongly tagged
as ``\nue at source'' is 1.3\%, i.e. the effective livetime of the
detector is about 99\%.  It sets more stringent limit than pile-up to
the scalability of this technique up to the Superbeams: already an
order of magnitude increase of $ N_\pi$ (100~kW for a 14\% duty cycle)
would bring the livetime below 70\%. To cope with Superbeam powers,
time resolutions of the order of a few hundreds of ps would be needed
both at the tracker and at the neutrino detector.

Finally, it is worth noting that Eq.~\ref{eq:acc} demonstrates
quantitatively what stated in Sec.~\ref{sec:setup}, i.e. the need of a
high speed tracker in front of the calorimeter units with time
resolution of $\sim$1~ns. Relying on the calorimetric measurement only
would increase the rate of accidental coincidences by about one order
of magnitude.
 
The second source of accidentals comes from events without positrons
in the final state misidentified as ``\nue at source'' due to finite
detector resolution ($\mu \rightarrow e$ or $\pi \rightarrow e$
misidentification). For the bulk of muons in the 2-body decay of
$\pi^+$, the misidentification probability is $5.4 \times 10^{-3}$,
resulting into a probability of a fake veto of 0.3\%.  However,
two-body hadronic decays of $K^+$ can give a relatively large rate of
false tagging.  This is due to the fact that pions are produced at
large angles in association with photons from $\pi^0$ decay. Since the
$\pi \rightarrow e$ misidentification probability is large
($\sim$25\%, see Fig.~\ref{fig:edep}) the accidental rate of $K^+
\rightarrow \pi^+ \pi^0$ (1.4\%) is comparable with the rate due to
genuine \nue sources. Clearly, the poor $\pi/e$ rejection capability
in the tagger is due to the fact that the longitudinal development of
the hadronic versus e.m. shower is not exploited for $\pi/e$
separation (see Sec.~\ref{sec:beam_dump}). Other sources of background
are listed in Tab~\ref{tab:background} and do not exceed 1\%.

\begin{table}
\centering
\begin{tabular} { c c c c }
\hline
 Source  & BR   & Misid & fake tag prob \\
\hline
$\pi^+ \rightarrow \mu^+ \nu_\mu$ & 100\% & $\mu \rightarrow e$ misid. & 0.3\% \\
$\mu^+ \rightarrow e^+ {\bar \nu_\mu} \nu_\mu$ & DIF & genuine $e^+$ & 0.1\% \\
$K^+ \rightarrow  \mu^+ \nu_\mu$  & 63.5\% & $\mu \rightarrow e$ misid. & $<$0.1\% \\
$K^+ \rightarrow  \pi^+ \pi^0$  & 20.7\% & $\pi \rightarrow e$ misid. & 1.4\% \\
$K^+ \rightarrow  \pi^+ \pi^+ \pi^-$  & 5.6\% & $\pi \rightarrow e$ misid. & 0.7\% \\
$K^+ \rightarrow  \pi^0 e^+ \nu_e$  & 5.1\% & genuine $e^+$ & 1.1\% \\
$K^+ \rightarrow  \pi^0 \mu^+ \nu_\mu$  & 3.3\% & $\mu \rightarrow e$ misid. & $<$0.1\% \\
$K^+ \rightarrow  \pi^+ \pi^0 \pi^0$  & 1.7\% & $\pi \rightarrow e$ misid. & $<$0.1\% \\
\hline
\end{tabular}
\label{tab:background}
\caption{Sources of background and their contribution to the probability of false tagging. }
\end{table}

\section{Performance of an instrumented beam dump}
\label{sec:beam_dump}

\begin{figure}
\centering \epsfig{file=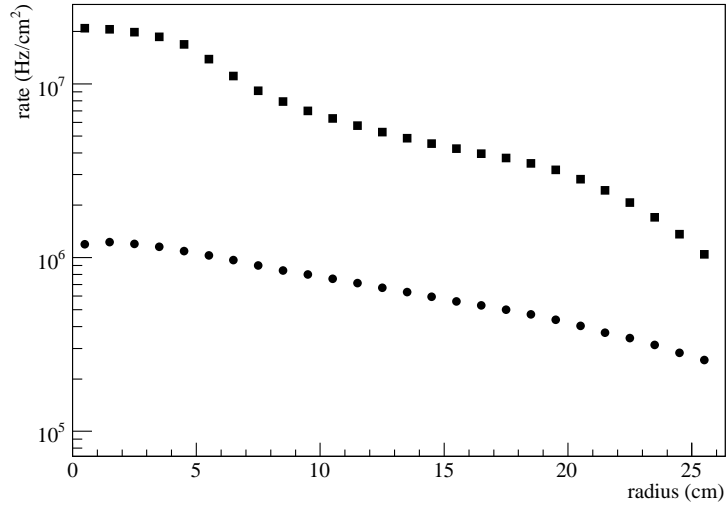,width=0.7\textwidth}
\caption{Charged particle rate (squares) and muon rate (dots) at the
last module for $R<26$~cm (instrumented beam dump). }
\label{fig:rate_dump}
\end{figure}

As shown in Sec.~\ref{sec:performance}, the performance of the tagging
system are excellent for the suppression of the \nue from \kthree
while a significant fraction of the positrons produced by the
decay-in-flight of muons impinges upon the hadron dump at the end of
the decay tunnel with a radius smaller than the inner radius of the
last tagging station. Fig.~\ref{fig:rate_dump} shows the particle rate
at the last tagging module at radii smaller than $R_{min}=26$~cm. For
the benchmark beamline, the rates do not exceed 20~MHz/cm$^2$, so that
it would be conceivable to instrument also the dump with a fast
tracker followed by a FCAL unit. Three issues, however, have to be
addressed, which in principle limit the tagging performance of the
dump and its scalability to higher powers. The first is related to the
integrated dose.  Unlike the scraping modules, the dump is crossed by
the bulk of undecayed pions and by secondary protons that deposit an
average energy fraction $\epsilon_\pi$ and $\epsilon_p$ of their
initial energy. The value of $\epsilon_{\pi,p}$ has been computed by
simulation and it turns out to be $\epsilon_\pi = 5.2 \pm 0.5$\% for
pions and $\epsilon_p = 4.0 \pm 0.5$\% for protons in 10~$X_0$ of
Pb. In the benchmark beamline the proton yield at the dump is large
and comparable with the one of the undecayed pions: for each
extraction $1.5 \times 10^{10}$ protons are distributed at the
entrance of the decay tunnel in the $10 \times 10$~cm$^2$ surface.  On
the other hand, if the beamline is operated in antineutrino mode
($\pi^-$ selection after the primary target), the corresponding
antiproton contribution is below the one of $K^-$.  Neglecting the
small contribution of undecayed $K^+$, the yearly integrated dose is,
therefore
\be \frac{   N_\pi \left[ (1-f) \epsilon_\pi E_\pi  + f E_{mip} \right] + N_p \epsilon_{p} E_p } {M}
\simeq 1.3\times 10^{-2} \ \mathrm{Gy/spill}
\label{eq:dose}
\ee
which corresponds to 25~kGy for a module weight of $M=115$~kg
(10~$X_0$ of Pb) and $10^{20}$~pot. 
In Eq.~\ref{eq:dose} $f$
represents the fraction of decayed pions (15\%), $E_\pi$ and $E_p$ the
mean pion and proton energy and $E_{mip}$ the energy released by a
muon in 10~$X_0$. The integrated dose is, therefore, much below the
safety operation limit of FCAL (5~MGy/y~\cite{atlas}).

\begin{figure}
\centering
\epsfig{file=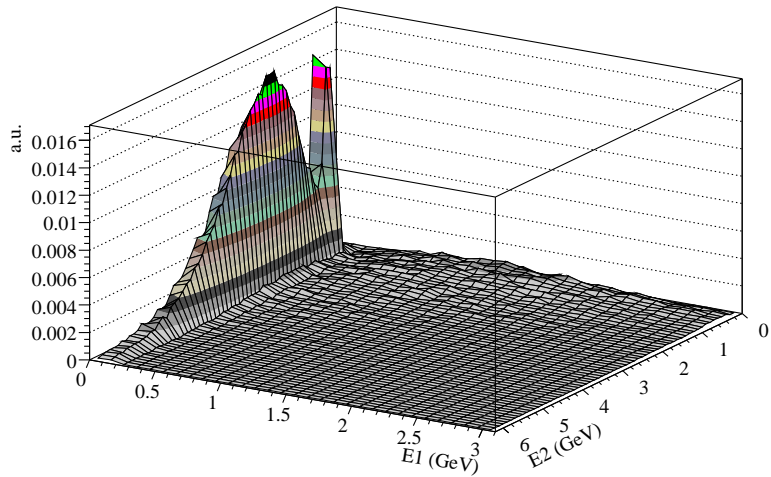,width=0.7\textwidth}
\caption{Energy deposition in the first 10~$X_0$ ($E1$) and in the
subsequent 3~$\lambda_I$ ($E2$) along a cylinder of $R_c=2$~cm
centered at the impact point of undecayed pions in the benchmark
beamline.}
\label{fig:selection}
\end{figure}

Once more, the issue that poses the main impediment to installing an
instrumented dump at higher beam powers is the rate of false tags from
random coincidences. Here, the number of
accidentals at the far detector per neutrino interaction  scales as:
\be 
\left[ N_\pi (1-f) \epsilon_{\pi \rightarrow e} + N_p \epsilon_{p \rightarrow e} \right] \cdot \Delta t
\simeq 23 \cdot \epsilon_{\pi \rightarrow e} 
\label{eq:acc_dump}
\ee
with $\Delta t \simeq 1$~ns and $\epsilon_{\pi \rightarrow e}$
($\epsilon_{p \rightarrow e}$)
representing the fraction of pions (protons) identified as electrons. In this
case, even at low power (benchmark beamline), the $\pi/e$
misidentification rate must be kept well below the value measured at
the tagger (see Sec.~\ref{sec:accidentals}). This is quite an ambitious
task, since the energy of the pions is much larger than the positron
spectrum in DIF events and, in order to reduce pile-up, only a small
area around the impact point can be used to collect the deposited
energy.  In this case, better performance can be obtained aligning two
FCAL modules of different thickness: the first one (FCAL1) is a
standard 10~$X_0$ tagger module, which is followed by a second
hadronic module FCAL2 of about three interaction lengths (thickness:
51~cm).  This configuration is quite similar to the one employed in
the ATLAS forward region~\cite{atlas,atlas2}. In this case, a more
powerful $\pi/e$ and $p/e$ separation can be achieved selecting events
in the $E1$ versus $E2$ plane (see Fig.~\ref{fig:selection}), $E1$ and
$E2$ being the visible energy deposited at FCAL1 and FCAL2,
respectively.  Again, the energy resolution for positrons is dominated
by the sampling term of the calorimeter while the hadronic resolution
is dominated by the lateral leakage of the hadronic energy at
$R_c>2$~cm.  $\epsilon_{\pi \rightarrow e}$ has been computed from the
distribution of $E1$ and $E2$ for events selected with the cut of
Tab.~\ref{tab:result1}, third column, (``Dump''), i.e.  requiring a
visible energy deposit in FCAL1 greater than 1~GeV and a small deposit
($<$300~MeV) in FCAL2. Here, the misidentification probability
$\epsilon_{\pi \rightarrow e}$ drops to 3\%. This value is still too
high for the benchmark beamline, especially when neutrinos are
produced and the proton contamination has to be accounted for (see
Eq.~\ref{eq:acc_dump}).

Finally, the tagging performance of the dump is challenged by event
pile-up, which is in fact entangled with the evaluation of
$\epsilon_{\pi \rightarrow e}$. For the benchmark beamline, even
neglecting the proton contribution, the pion density at the dump is $
N_\pi (1-f)/ \pi R_{min}^2 \simeq 4$~MHz/cm$^2$. So, for an effective
integration window $\Delta T_{cal}$ of 25~ns, the pile-up rate PR is
\be 
\frac{ N_\pi (1-f) }{ \pi R_{min}^2} \cdot \pi R_c^2 \cdot \Delta T_{cal} \ \ ;
\ee 
it corresponds to 1.2 event for $R_c=2$~cm. Once accounting for
pile-up, the $\pi \rightarrow e$ misidentification probability grows
up to $\sim 10$\% at constant detection efficiency for DIF
positrons. As a consequence, even if further suppression of the pion
background might be achieved exploiting the transverse shower profile
or a better longitudinal segmentation, the instrumented beam dump can
be fruitfully employed only for moderate beam powers ($\sim 10^{9}$
$\pi^+$ per extraction) or, equivalently, for time resolutions below
1~ns. At larger beam powers, instead of exploiting the instrumentation
of the dump, it is more convenient to reduce the angular spread at the
entrance of the decay tunnel (3~mrad in the present case) and,
therefore, increase the geometrical acceptance of the scraping taggers
at lower $R_{min}$.

\section{Conclusions}
\label{sec:conclusions}

In the last decade, outstanding progresses have been a\-chie\-ved in the
development of fast, radiation hard detectors for tracking and
particle identification. Such efforts have been motivated by
challenging requests from modern experiments at colliders - firstly
the LHC - and from experiments in the field of rare kaon decays. In
this paper, we discussed an application of these technologies aimed at
tagging $\nue$ neutrinos in beams originating from the decay-in-flight
of charged pions and based on scraping of secondary and tertiary
leptons along the decay tunnel. For a specific beam configuration of
moderate power ($\sim$10~kW for a 14\% duty cycle), we have shown that
the tagging system can achieve a suppression of 86\% of the \nue
background from \kthree decays and of about 60\% of the \nue from the
decay-in-flight (DIF) of $\mu^+$. This setup can also be employed for
beams of larger power (up to $\sim 100$~kW at the same duty cycle)
without significant loss in performance. At beamlines of a few kW
power, the tagging efficiencies for DIF \nue can be further improved
(+16\%) by an additional instrumented beam dump located at the end of
the decay tunnel; at larger powers the use of the instrumented dump is
limited by the rate of accidentals due to $\pi \rightarrow e$ and $p
\rightarrow e$ misidentification.

\section*{Acknowledgments}
We wish to express our gratitude to U. Dore, P. Loverre, M. Mezzetto
and F. Ronga for many interesting discussions and a very careful
reading of the manuscript. We are grateful to A.~Blondel and
R.~Steerenberg for useful information on the CERN-PS. A special thank
to R.~Felici, whose suggestions speeded up the completion of this
work.


\end{document}